# Assessing Data Center Hosting Capacity Considering Electromagnetic Transient Dynamics

Fan Jiang
*Department of Electrical and Computer Engineering*
*University of Houston*
Houston, TX, USA
fjiang6@uh.edu

Xingpeng Li
*Department of Electrical and Computer Engineering*
*University of Houston*
Houston, TX, USA
xli83@central.uh.edu

***Abstract*—Hosting capacity is commonly assessed with steady-state optimization, which leaves the dynamic response of the data center and the grid at the resulting operating point untested. This paper assesses candidate data center capacities by coupling a unit commitment (UC) model with electromagnetic transient (EMT) simulation. During the capacity search, the UC model determines the economic dispatch for each tested data center capacity, and the resulting operating points are tested for settling with the generators committed online by the UC model and after a selected synchronous generator outage. Three nested capacities are obtained: the optimization-based steady-state capacity, the dynamic pre-outage capacity, and the post-outage dynamic capacity, each bounded by the previous one. On a test system with a detailed data center model, both dynamic capacities fall well below the optimization capacity, and the ranking of connection locations changes with the assessment stage and with the outage. In the most severe outage case at each candidate bus, the capacity is about 30% below the steady-state capacity. Thus, the dynamic response of the data center and the grid should be part of the hosting capacity assessment, and the comparison of connection locations should be made on the post-outage dynamic capacity.**

## I. Introduction

Data centers support essential digital services and artificial intelligence (AI), and their increasing electricity demand makes reliable grid integration an important planning concern [1], [2]. Hosting capacity specifies the maximum demand that can be accommodated at a connection location under stated network conditions and assessment criteria. For data centers, it provides a basis for evaluating connection requests and comparing candidate locations. What a reported capacity guarantees depends on how it is obtained: a capacity that satisfies a steady-state model does not show whether the system can ride through a disturbance at the associated operating point.

Most studies of data center hosting capacity rely on steady-state network models. Transmission hosting capacity maps proposed in [3] show the load that each location can accommodate. Vafamehr et al. [4] jointly planned data center locations, capacities, and data routes with network feasibility and generation dispatch under uncertainty. Kim et al. [5] optimized the siting and dispatch of flexible data center loads, while Xiao et al. [6] coordinated siting with renewable generation dispatch and transmission alternatives. Qiu et al. [7] characterized hosting capacity for distributed generative AI data centers using spatiotemporal feasible regions. Manassero et al. [8] combined data center capacity assessment and location optimization with equipment outage screening based on steady-state voltage and current limits. These studies quantify the capacity that the network admits, but their checks do not include the dynamic performance of the data center, which can reduce the capacity that is actually acceptable.

Data center dynamics have been shown to affect the power system. J. Sun et al. [9] showed that transfer to an uninterruptible power supply (UPS) during a voltage sag keeps the servers supplied while the grid import drops to zero, so the grid sees a sudden loss of load. Ko and Zhu [10] examined how AI workload spectra, facility size, and location influence system oscillations. Chaudhary et al. [11] used real-time electromagnetic transient (EMT) simulation to show that spatially correlated data center loads enlarge the aggregate disturbance, deepen frequency dips, and reduce voltage stability margins. These findings show that the response of a data center to a disturbance is set by its converters and protection controls, which a steady-state load model cannot represent. A hosting capacity assessment should therefore include an EMT model to capture these data center dynamics.

Data center dynamics have also been incorporated into capacity assessment and operational optimization. Samir et al. [12] used measured graphics processing unit power profiles to find the capacity of AI training data centers that satisfies voltage and frequency limits, and then allocated it among candidate buses. However, the capacity is not tied to an economic operating point, and the disturbances are the power swings of the data center itself rather than grid contingencies. Ren et al. [13] incorporated data center fast frequency response into unit commitment (UC). Their model has the economic schedule but no capacity decision. R. Sun et al. [14] used data-driven inverse optimization to reduce the dimensionality of constraints in data center load models and developed a scheduling strategy based on the reduced model for ancillary services.

Although dynamic hosting capacity and frequency-constrained scheduling of data centers have been investigated, the link between candidate capacity, its economic operating point, and the detailed response to generator outages remains insufficiently explored. This paper couples UC with EMT simulation to assess each data center capacity during the capacity search procedure at its corresponding economic operating points. The contributions are as follows:

- An assessment framework is proposed. The capacity search procedure links the operating point from the optimization model with the dynamic response from EMT simulations.
- The framework has a pre-outage stage and a post-outage stage and yields three nested capacities: the optimization-based steady-state capacity, pre-outage dynamic capacity, and post-outage dynamic capacity.
- A case study on the modified IEEE 9-bus system shows that both dynamic capacities fall well below the steady-state capacity and that the ranking of the candidate buses changes with the assessment stage and the outage.

The rest of the paper is organized as follows. Section II formulates the steady-state hosting capacity model, Section III presents the dynamic assessment, Section IV reports the case study, and Section V concludes.

## II. Steady-State Hosting Capacity Model

The steady-state hosting capacity model gives the data center hosting capacity under steady-state constraints alone. It is formulated as a UC model whose objective is to maximize the data center capacity.

Given a data center location $d$, the hosting capacity model maximizes the data center capacity as

$$C_d^{UC} = \max C_d \tag{1}$$

where $C_d$ is the data center capacity at location $d$, and $C_d^{UC}$ is the resulting optimization capacity.

The data center is modeled as a constant load equal to its capacity throughout the horizon, whereas the other loads follow the load multiplier. The data center is added to the original demand at bus $d$, so the scheduled demand $D_{bt}$ is

$$D_{bt} = \begin{cases} \lambda_t \overline{D}_b + C_d, b = d \\ \lambda_t \overline{D}_b, b \neq d \end{cases}, \forall b \in \mathcal{N}, t \in \mathcal{T} \tag{2}$$

where $\overline{D}_b$ is the nominal demand at bus $b$, $\lambda_t$ is the load multiplier of interval $t$, $\mathcal{N}$ is the set of buses, and $\mathcal{T}$ is the set of intervals.

The constraints below apply to every synchronous generator (SG), bus, branch, and interval. For each SG and interval, output and steady-state reserve satisfy

$$P_g^{\min} u_{gt} \leq P_{gt}, \forall g \in \mathcal{G}, t \in \mathcal{T} \tag{3}$$

$$P_{gt} + R_{gt} \leq P_g^{\max} u_{gt}, \forall g \in \mathcal{G}, t \in \mathcal{T} \tag{4}$$

$$0 \leq R_{gt} \leq R_g^{\text{res}} u_{gt}, \forall g \in \mathcal{G}, t \in \mathcal{T} \tag{5}$$

$$\sum\nolimits_{h \in \mathcal{G}, h \neq g} R_{ht} \geq P_{gt}, \forall g \in \mathcal{G}, t \in \mathcal{T} \tag{6}$$

where $P_{gt}$, $R_{gt}$, and $u_{gt}$ are the output, the reserve, and the online state of SG $g$ in interval $t$; $P_g^{\min}$, $P_g^{\max}$, and $R_g^{\text{res}}$ are the minimum output, maximum output, and reserve capability of SG $g$; and $\mathcal{G}$ is the set of SGs.

The ramp limits are

$$P_{gt} - P_{g,\pi(t)} \leq R_g^{\text{ramp}} u_{g,\pi(t)} + R_g^{\text{start}} v_{gt} \tag{7}$$

$$P_{g,\pi(t)} - P_{gt} \leq R_g^{\text{ramp}} u_{gt} + R_g^{\text{start}} (v_{gt} - u_{gt} + u_{g,\pi(t)}) \tag{8}$$

$$\pi(t) = \begin{cases} t-1, t > 1 \\ T, \ t = 1 \end{cases}, \forall \, t \in \mathcal{T} \tag{9}$$

where $R_g^{\text{ramp}}$ is the normal ramp allowance per interval; $R_g^{\text{start}}$ is the startup and shutdown allowance; $v_{gt}$ is the startup state of SG $g$ in interval $t$; and $T$ is the scalar total number of intervals.

The generator minimum on/off time constraints are

$$\sum\nolimits_{s \in \mathcal{W}_{gt}^-} v_{gs} \leq u_{gt} \tag{10}$$

$$\sum\nolimits_{s \in \mathcal{W}_{gt}^+} v_{gs} \leq 1 - u_{gt} \tag{11}$$

where $\mathcal{W}_{gt}^-$ is the set of intervals whose startups require SG $g$ to be online in interval $t$, namely interval $t$ and the preceding intervals, with a window length equal to the minimum online duration of SG $g$; $\mathcal{W}_{gt}^+$ is the set of intervals whose startups require SG $g$ to be offline in interval $t$, namely the intervals after $t$ with a window length equal to the minimum offline duration of SG $g$. Both time windows wrap around the scheduling horizon, consistent with (9).

The startup state is linked to the online states, and a startup is allowed only when the SG is offline in the preceding interval, so that

$$u_{gt} - u_{g,\pi(t)} \leq v_{gt} \tag{12}$$

$$v_{gt} \leq 1 - u_{g,\pi(t)} \tag{13}$$

$$u_{gt}, v_{gt} \in \{0,1\} \tag{14}$$

The network is represented by a DC power flow. The nodal active power balance is

$$\sum\nolimits_{g \in \mathcal{G}_b} P_{gt} + \sum\nolimits_{\ell \in \mathcal{L}_b^{\text{in}}} F_{\ell t} - \sum\nolimits_{\ell \in \mathcal{L}_b^{\text{out}}} F_{\ell t} = D_{bt} \tag{15}$$

where $\mathcal{G}_b$, $\mathcal{L}_b^{\text{in}}$, and $\mathcal{L}_b^{\text{out}}$ are the sets of SGs, incoming branches, and outgoing branches at bus $b$.

The branch flow, its limit, and the angle reference are

$$F_{\ell t} = \frac{\theta_{o(\ell),t} - \theta_{r(\ell),t}}{x_\ell} \tag{16}$$

$$-F_\ell^{\max} \leq F_{\ell t} \leq F_\ell^{\max} \tag{17}$$

$$\theta_{b_{\text{ref}},t} = 0 \tag{18}$$

where $o(\ell)$ and $r(\ell)$ are the sending and receiving buses of branch $\ell$, $x_\ell$ is its reactance, $F_\ell^{\max}$ is its flow limit, and $b_{\text{ref}}$ is the reference bus.

## III. Dynamic Hosting Capacity Assessment

This section proposes a data center dynamic hosting capacity assessment framework to obtain the pre-outage and post-outage capacities.

### A. Assessment Procedure

The procedure has two stages that share one evaluation loop, as shown in Fig. 1. The optimization model of Section II gives $C_d^{\text{UC}}$, which enters Stage 1, the pre-outage assessment, as the initial upper bound $U$; its output $C_d^{\text{pre}}$ enters Stage 2, the post-outage assessment, which returns $C_{dk}^{post}$.

Both stages call the capacity search and EMT evaluation block, which tests a candidate data center capacity $C$ at the selected bus within an interval $[L, U]$. The upper bound $U$ uses the preceding stage's output. The lower bound $L$ starts at zero and records the largest accepted candidate. A fixed-capacity UC dispatch initializes the EMT model, which runs with the scheduled SGs in Stage 1 and the selected SG outage in Stage 2. A candidate passes if the assessment criterion is met in the Stage 1 window or both Stage 2 windows. If the candidate passes, $L$ is updated to $C$; otherwise, $U$ is updated to $C$. The search ends when the bounds are one capacity step apart or the upper bound is accepted, and $L$, the largest accepted candidate found, is returned as the stage output, $C_d^{\text{pre}}$ in Stage 1 and $C_{dk}^{post}$ in Stage 2. If none passes, no accepted capacity is reported. Stage 2 is repeated for each SG outage, retaining the smallest capacity.

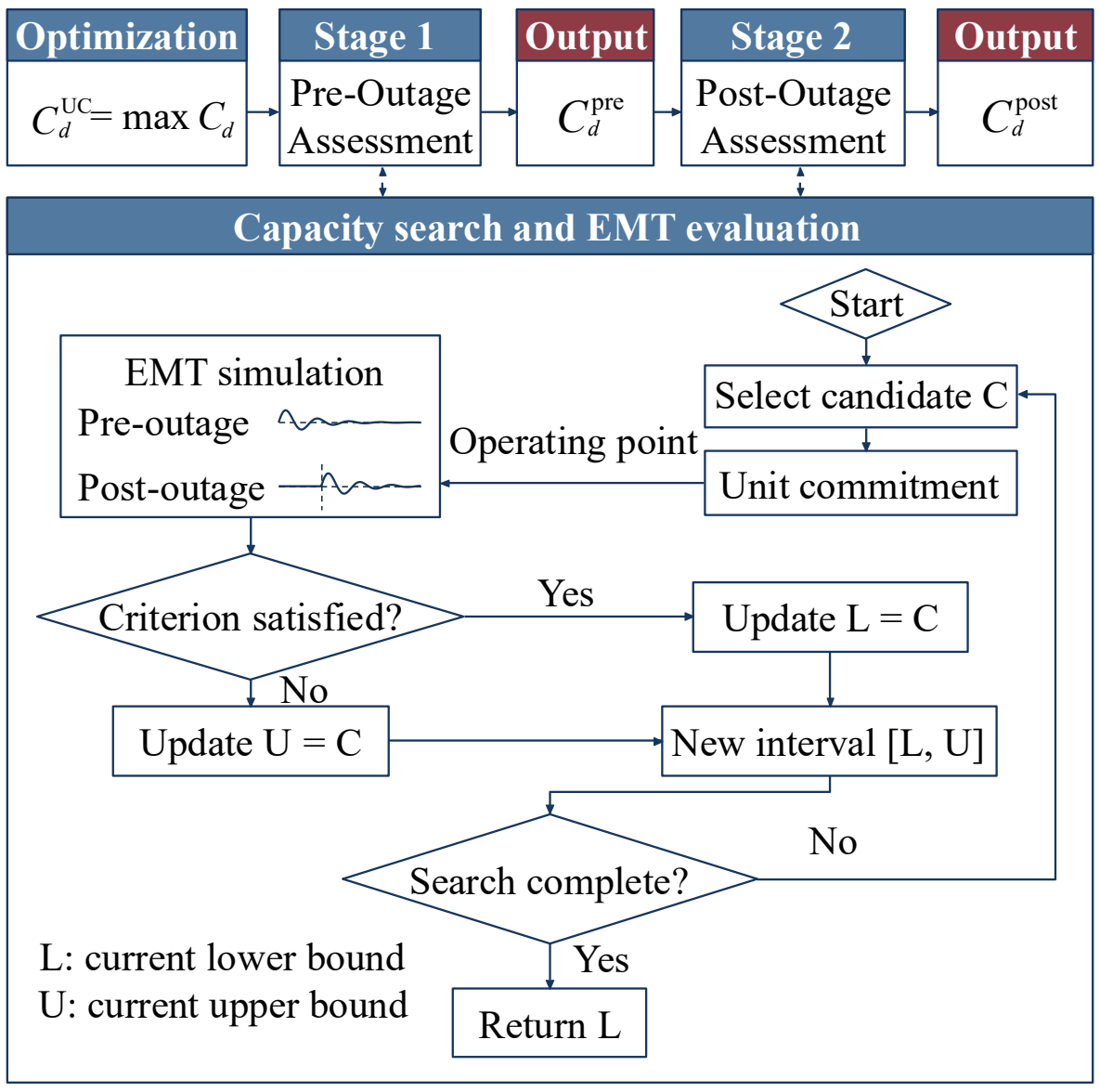

Fig. 1. Assessment procedure.

## B. Unit Commitment and EMT Models

For a candidate capacity $C$, the UC model is solved first in the capacity search and EMT evaluation block. The UC model keeps the constraints (2)–(18) of Section II, holds $C_d$ fixed at $C$, and replaces the objective by the operating cost

$$\min \sum_{t\in\mathcal{T}} \sum_{g\in\mathcal{G}} (a_g P_{gt} + b_g u_{gt} + s_g v_{gt}) \qquad (19)$$

where $a_g$, $b_g$, and $s_g$ are the output cost per interval, the online cost per interval, and the startup cost of SG $g$; the interval duration is included in $a_g$.

The two models share the same constraints and differ only in the objective. The hosting capacity model returns the maximum capacity for a specific location, while the UC model returns the commitment and generator outputs of every interval, and these values form the operating point passed to the EMT model. Specifically, in the EMT model, the SG active power setpoints are set to the UC values of the selected interval, the other loads are scaled by the load multiplier of that interval, and the data center model is scaled to the candidate capacity.

The EMT model represents the network, the SGs with their controls, and the data center at the candidate bus. The data center model includes the grid side rectifier, the inverter that supplies the computing load, the DC link with the battery and its bidirectional converter, the cooling load, and the transfer, bypass, protection, and reconnection logic of the UPS [15]. In the UPS control, the rectifier is blocked when the voltage at the data center AC bus stays below a threshold for a confirmation time and is re-enabled after a delay once the condition clears. The battery discharges while the rectifier is blocked and returns to charge mode after a further delay once the rectifier is re-enabled. The detailed configurations and control logic are in [15].

## C. Stage 1: Pre-Outage Assessment

The pre-outage assessment searches for the largest candidate capacity at which the system settles over one final observation window at its own economic operating point from the UC model without any SG outage.

The optimization capacity $C_d^{\mathrm{UC}}$ obtained from the hosting capacity model is passed to the capacity search and EMT evaluation block as the upper bound for Stage 1. The block returns the pre-outage hosting capacity $C_d^{\mathrm{pre}}$, which is passed to the same block as the upper bound for Stage 2.

## D. Stage 2: Post-Outage Assessment

The post-outage assessment applies the search with a selected SG outage $k$, starting from $C_d^{\mathrm{pre}}$ as its upper bound and zero as its lower bound. For each candidate data center capacity, the UC model determines the economic dispatch used to initialize the EMT simulation. During the simulation, the selected SG is disconnected, and the surviving SGs, the network, and the data center controls respond on their own. The assessment criterion is applied to the online SGs in a pre-outage observation window and to the surviving SGs in a final post-outage observation window. The largest candidate found by the search that passes both checks is output as $C_{dk}^{\mathrm{post}}$.

For each location $d$, (20) retains the smallest of the single SG outage results as the post-outage capacity $C_d^{\mathrm{post}}$. The outages are searched separately for all SGs, and the minimum is the post-outage hosting capacity $C_d^{\mathrm{post}}$ returned by this stage.

$$C_d^{\mathrm{post}} = \min_{k\in\mathcal{G}} C_{dk}^{\mathrm{post}} \qquad (20)$$

## E. Assessment Criterion

The EMT evaluation applies the same assessment criterion to each observation window. For signal samples $y_j = y(\tau_j)$ with sample index set $W$ in that window, the range $A_y$ and the absolute fitted slope $B_y$ are defined as

$$A_y(W) = \max_{j\in W} y_j - \min_{j\in W} y_j \qquad (21)$$

$$B_y(W) = \left| \frac{\sum_{j\in W} (\tau_j-\bar{\tau})(y_j-\bar{y})}{\sum_{j\in W} (\tau_j-\bar{\tau})^2} \right| \qquad (22)$$

where $j$ indexes samples, $\tau_j$ is time, and bars denote sample means over $W$.

For every online SG, the frequency $f_g$ and the measured active power $p_g$ must satisfy

$$\begin{cases} A_{f_g} \le \varepsilon_f \\ B_{f_g} \le \varepsilon_{\dot{f}} \\ A_{p_g} \le \varepsilon_p \\ B_{p_g} \le \varepsilon_{\dot{p}} \end{cases} \qquad (23)$$

where $\varepsilon_f$ and $\varepsilon_p$ bound the ranges of frequency and power, and $\varepsilon_{\dot{f}}$ and $\varepsilon_{\dot{p}}$ bound their slopes.

In Stage 1, the criterion is applied to the final observation window of the no-outage simulation, whereas in Stage 2 it is applied to observation windows both before and after the outage. The criterion judges only whether the response settles. The circuit breaker and rectifier blocking actions follow the data center model.

With the assessment criterion defined, the two stages in Fig. 1 can be carried out for any connection bus and any selected outage. The next section applies the procedure to a test system and examines the resulting capacities and the responses behind them.

## IV. Case Study

### A. System and Simulation Settings

The proposed assessment framework is applied to the modified IEEE 9-bus system shown in Fig. 2. One data center is added at bus 5, 6, or 8 in separate studies. The UC model is developed in Python with Pyomo according to the topology in Fig. 2. The UC horizon is 24 hourly intervals. The load profile follows the ERCOT system wide hourly load of Aug. 17, 2025 [16]. The EMT simulation is initialized at the operating point of the peak load interval, hour 18. The capacity step is 1 MW, so the candidates are integer capacities, and the search ends when the bounds are 1 MW apart or the upper bound is accepted. Each observation window is 5 s. All three SGs are committed online in the evaluated schedules, and the outages of all three SGs, G1, G2, and G3, are selected in Stage 2.

The EMT model is built in PSCAD and the data center model of [15] is connected to the candidate bus through transformers. The tolerances in (23) are 0.020 Hz, 0.005 Hz/s, 0.050 per unit, and 0.010 per unit/s on a 100 MVA base.

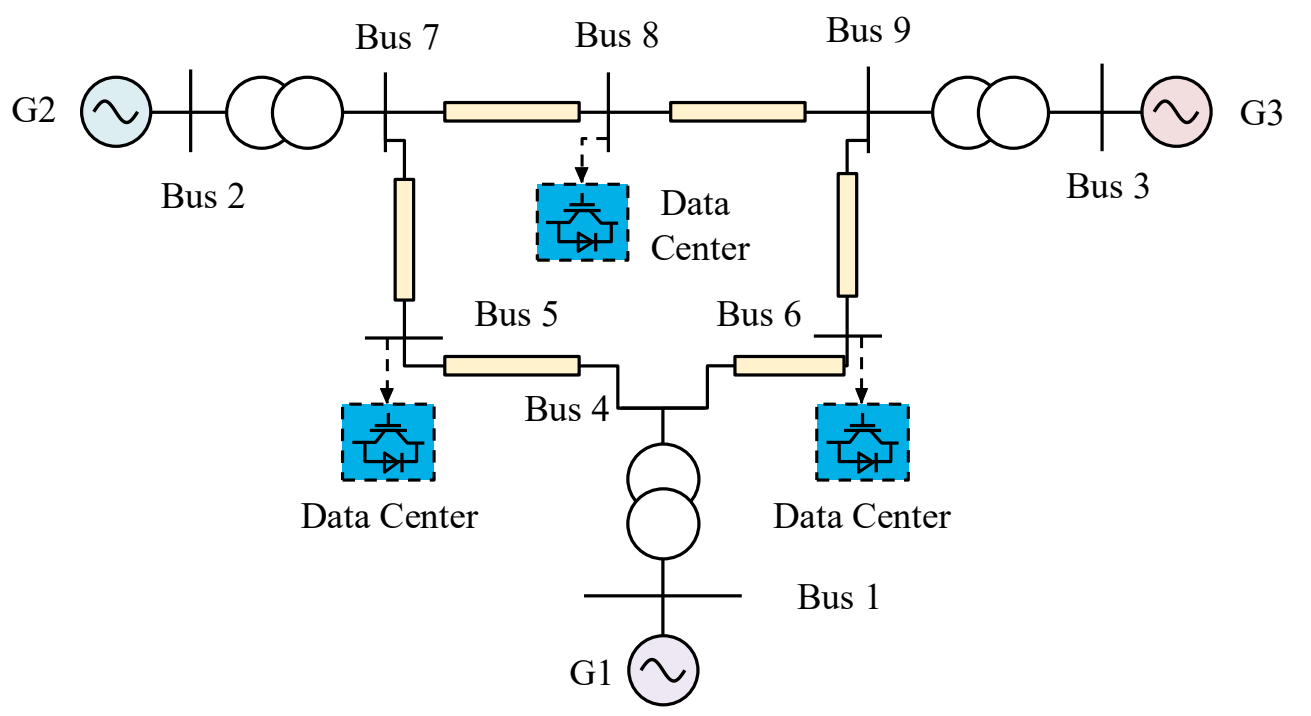

Fig. 2. Modified IEEE 9-bus system with the candidate data center buses.

### B. Capacity Comparison and Operating Point Dependence

Fig. 3 compares the optimization capacity $C_d^{\mathrm{UC}}$ with the pre-outage capacity $C_d^{\mathrm{pre}}$ and the post-outage capacities $C_{dk}^{post}$ for the G1, G2, and G3 outages when the data center is connected to bus 5, 6, or 8.

The gray bars in Fig. 3 are the optimization capacities $C_d^{\mathrm{UC}}$ from the UC model alone, approximately 212 MW at each bus.

Pre-outage capacities are lower than optimization capacities at every bus. The blue bars give the pre-outage capacities $C_d^{\mathrm{pre}}$ of 150, 192, and 169 MW, which are 29.1%, 9.2%, and 20.1% below the optimization capacities. The corresponding reductions are about 62 MW, 20 MW, and 43 MW at buses 5, 6, and 8, respectively. UC includes network and intertemporal constraints but omits the converter and SG dynamics evaluated in EMT.

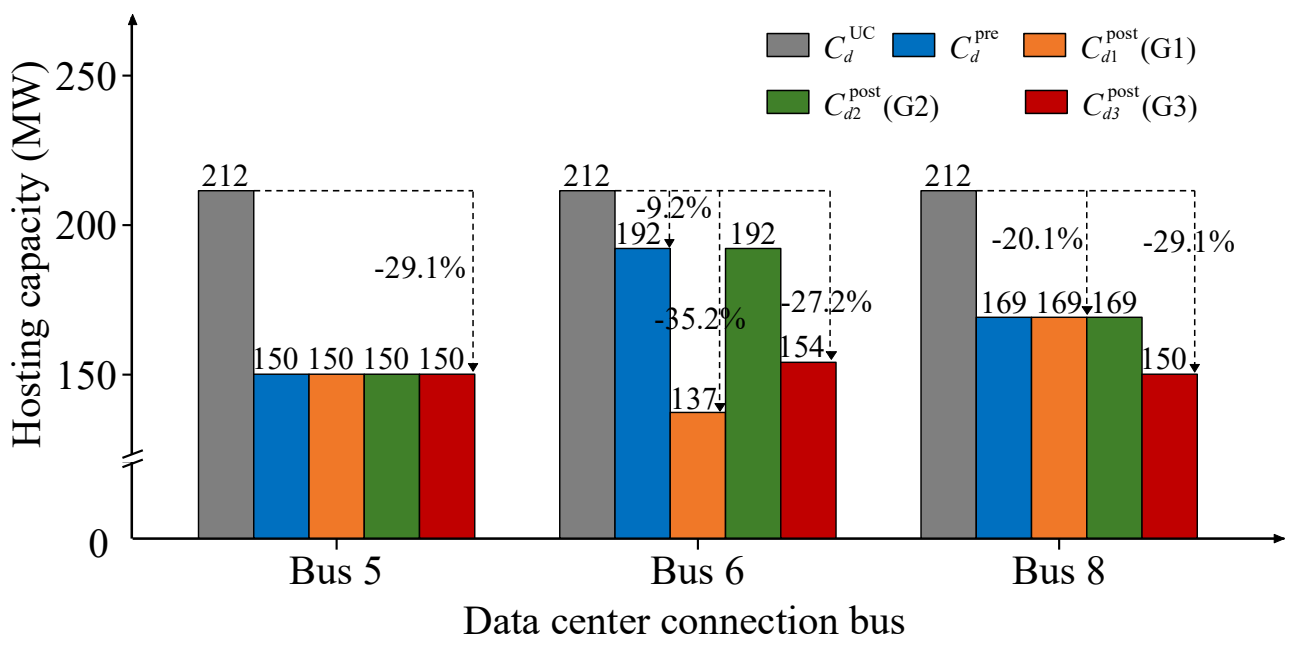

Fig. 3. Hosting capacities under different scenarios.

An SG outage can lower hosting capacity further. The post-outage capacities $C_{dk}^{post}$ are 9.2%, 29.1%, 35.2% below $C_d^{\mathrm{UC}}$, and several outages leave the pre-outage capacity unchanged. The reserve constraint (6) requires sufficient reserve from the remaining online SGs to cover the output of the lost generator, but does not model the dynamic deployment of that reserve. Pre-outage acceptance therefore does not guarantee post-outage acceptance.

The minimum post-outage capacities are about 30% below $C_d^{\mathrm{UC}}$. The final post-outage capacities are 150 MW, 137 MW, and 150 MW at buses 5, 6, and 8, respectively, according to (20). Compared with the pre-outage capacities, these values represent additional reductions of 0%, 28.6%, and 11.2%, respectively.

The N−1 criterion requires the system to recover following any single SG outage considered. The data center hosting capacity that satisfies this criterion depends on the connection bus. At bus 5, all three SG outages yield the same post-outage hosting capacity. At buses 6 and 8, the G1 and G3 outages are the most restrictive, respectively.

The ranking of the candidate buses changes with the assessment stage. The optimization capacities are equal, whereas bus 6 ranks first in the pre-outage assessment, exceeding bus 5 by 42 MW. Bus 8 ranks first under the G1 outage, while bus 6 ranks first under the G2 and G3 outages. Buses 5 and 8 share the highest final post-outage capacity, exceeding bus 6 by 13 MW. The capacity spread across the three buses decreases from 42 MW in the pre-outage assessment to 13 MW in the final post-outage assessment.

In summary, the pre-outage assessment returns lower hosting capacities than the optimization model at every bus, and the minimum post-outage capacities can be lower still. The most restrictive outage depends on the bus, with final capacities about 30% below the optimization capacity, and the ranking of the buses changes from one stage to the next. The hosting capacity of a data center should therefore be assessed with its dynamic response included.

### C. Frequency Response During SG Outage

The additional capacity reductions following an SG outage are associated with the combined dynamic response of the generators, network, and data center. When an SG is lost, the system frequency normally falls because the demand exceeds the supply. A data center, however, is a converter-interfaced load that switches to backup mode to protect itself, so the transient after the outage can trigger its protection and remove a large block of demand from the grid, which changes the frequency response from that of a conventional load.

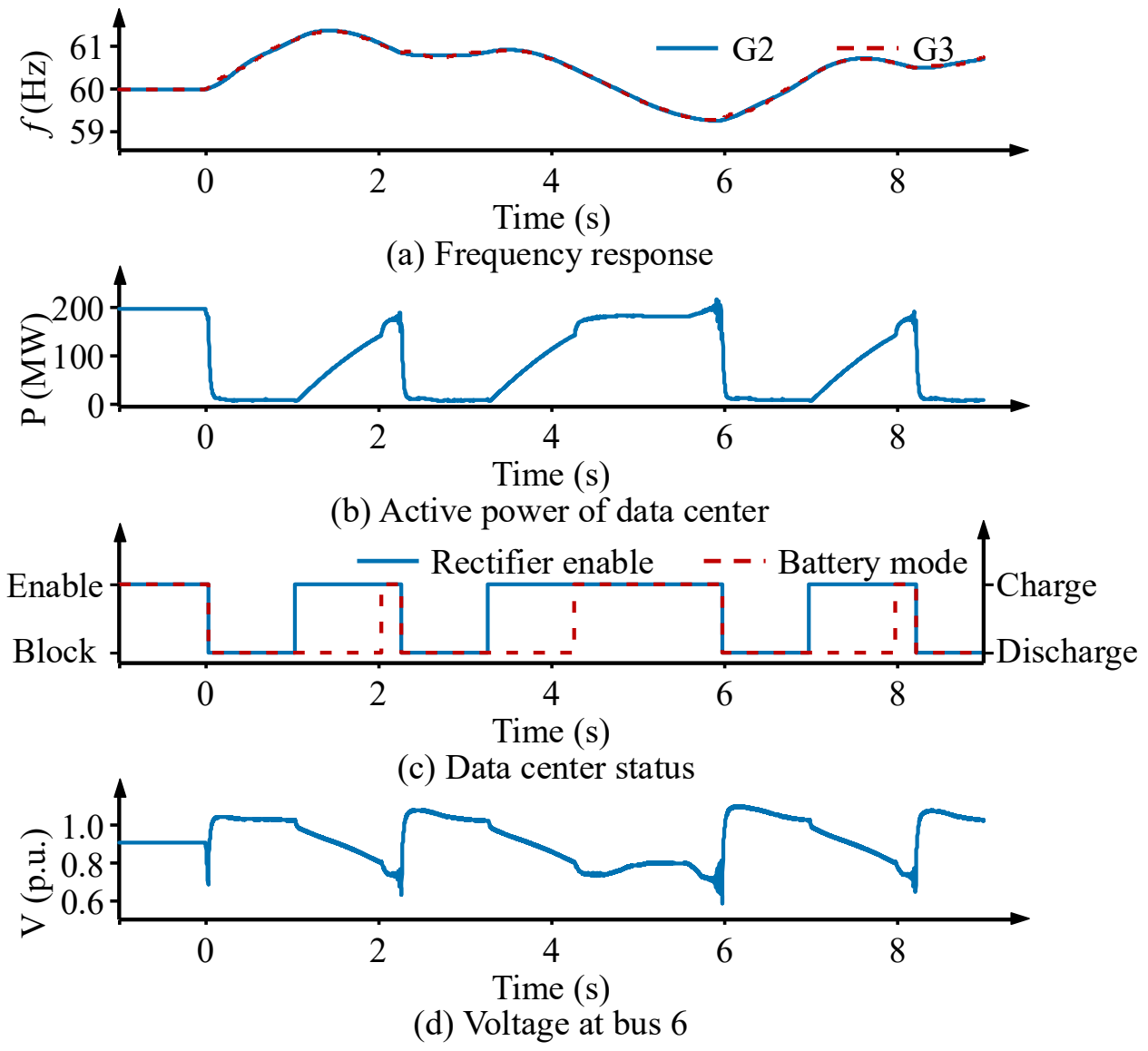

Fig. 4. Response at bus 6 following the G1 outage: (a) frequency response, (b) active power of the data center, (c) data center status, (d) bus voltage.

In Fig. 4, the data center is connected at bus 6, and the capacity is approximately 198 MW before the G1 outage. It is worth noting that this case fails the Stage 2 assessment criterion.

The G1 outage defines 0 s in Fig. 4. As shown in Fig. 4(a), the frequency rises after the outage instead of falling. Shortly after the outage, the data center blocks its rectifier in response to the voltage drop in Fig. 4(d), and the UPS battery control consequently switches to discharge mode, as shown in Fig. 4(c). The active power imported to the data center collapses to a small fraction of its pre-outage value, as shown in Fig. 4(b). This temporary demand reduction is far larger than the pre-outage output of G1, so the surviving SGs accelerate despite the loss of generation, and their frequencies rise well above nominal before falling.

The rectifier is re-enabled about 1 s after the outage, and the battery control returns to charge mode about 1 s later. The frequency then falls as the grid import recovers. Blocking and recovery continue to recur several times, and finally the criterion is not satisfied. These control actions shape the post-outage response and the capacity it supports, which the steady-state model cannot capture.

## V. Conclusion

This paper proposed a hosting capacity assessment framework that obtains the pre-outage and post-outage hosting capacities of a data center by coupling a UC model with EMT simulation. The optimization capacity from the UC model bounds a capacity search in which the UC model determines the economic dispatch for each candidate capacity and its response in the EMT model is checked against an assessment criterion. In the case study on a modified IEEE 9-bus system, both dynamic capacities fall well below the optimization capacity, the final post-outage capacities are about 30% below the optimization capacity, the ranking of the candidate buses changes with the assessment stage and the outage, and the rectifier blocking of the data center controls drives a frequency overshoot that the steady-state model cannot capture.

## Acknowledgement

This research was supported in part by the University of Houston Energy Transition Institute Seed Grant Program.